\documentclass[draft]{agujournal2019}
\usepackage{url} 
\usepackage{lineno}
\usepackage[inline]{trackchanges} 
\usepackage{soul}
\usepackage{amsmath}
\graphicspath{ {./} }

\usepackage{color}

\draftfalse

\journalname{Space Weather}

\begin{document}

%
%


\title{Prediction of solar wind speed by applying convolutional neural network to potential field source surface (PFSS) magnetograms}

%
%




\authors{Rong Lin\affil{1}, Zhekai Luo\affil{1}, Jiansen He\affil{1}, Lun Xie\affil{1}, Chuanpeng Hou\affil{1}, Shuwei Chen\affil{2}}

\affiliation{1}{School of Earth and Space Sciences, Peking University \\
Beijing, 100871, China}
\affiliation{2}{School of Artificial Intelligence, Nanjing University, Nanjing, 210023, China}





\correspondingauthor{Jiansen He}{jshept@pku.edu.cn}




\begin{keypoints}
\item A model based on convolutional neural network is trained to predict the solar wind speed upstream of the Earth.
\item This model uses four potential field source surface (PFSS) magnetograms three, four, five, and six days before the prediction as the input.
\item The model achieves a performance of CC=0.52 and RMSE=80.8km/s on the continuous test dataset.
\end{keypoints}

%
%

%
%


\begin{abstract}
An accurate solar wind speed model is important for space weather predictions, catastrophic event warnings, and other issues concerning solar wind – magnetosphere interaction. In this work, we construct a model based on convolutional neural network (CNN) and Potential Field Source Surface (PFSS) magnetograms, considering a solar wind source surface of $R_{\rm SS}=2.5R_\odot$, aiming to predict the solar wind speed at the Lagrange-1 (L1) point of the Sun-Earth system. The input of our model consists of four Potential Field Source Surface (PFSS) magnetograms at $R_{\rm SS}$, which are 7, 6, 5, and 4 days before the target epoch. Reduced magnetograms are used to promote the model's efficiency. We use the Global Oscillation Network Group (GONG) photospheric magnetograms and the potential field extrapolation model to generate PFSS magnetograms at the source surface. The model provides predictions of the continuous test dataset with an averaged correlation coefficient (CC) of 0.52 and a root mean square error (RMSE) of 80.8 km/s in an eight-fold validation training scheme with the time resolution of the data as small as one hour. The model also has the potential to forecast high speed streams of the solar wind, which can be quantified with a general threat score of 0.39.
\end{abstract}

\section*{Plain Language Summary}
The dynamic pressure of the solar wind is a crucial condition for solar activity to affect Earth's space weather. The strength of the solar wind's dynamic pressure depends on the solar wind's speed. Therefore, predicting solar wind velocity upstream of the Earth is one of the essential topics of space weather research. The distribution of the solar coronal magnetic field is an important factor in regulating the speed of the solar wind. In this work, we explore the mapping from the variation of the extrapolated solar coronal magnetic field at the source surface, where the open field lines are assumed to direct radially along with the nascent solar wind flow, to the variation of solar wind speed at the Lagrange-1 (L1) point of the Sun-Earth system. We use the photospheric magnetograms from the Global Oscillation Network Group (GONG) to extrapolate the coronal magnetic field up to the source surface. We construct a prediction model based on convolutional neural networks (CNN) with five hidden layers. The model proves that the mapping relation is reliable and has an advantage in predicting when the solar wind speed starts to grow.

%
%

\section{Introduction}
The solar wind, as a continuous supersonic plasma flow emanating from the solar corona, fills in the interplanetary space between the Sun and various planets, including the Earth, and contributes to the formation of the heliosphere. The solar wind carries interplanetary magnetic fields (IMFs) and interacts with the Earth's magnetic field, forming the geo-magnetopause, which separates the shocked solar wind from the Earth's magnetosphere.

The solar wind is highly structured and dynamic. According to its radial speed, it can be generally categorized into two kinds of streams: the high-speed stream (HSS) and the low-speed stream (LSS)\cite{schwenn1983average}. The origin of the high-speed stream has been widely recognized as coronal holes \cite{Wilcox1968largeScale, Altschuler1969, Hollweg2002}, while the low-speed stream has multiple types of source region \cite{Antiochos2011, abbo2016slow, SanchezDiaz2019}. The activity of the SW can significantly affect the near-Earth space. In the interplanetary space, the high-speed stream compresses the low-speed stream, forming stream interaction regions (SIRs) and corotating interaction regions (CIRs) \cite{Gosling1978,Smith1976}. Some CIRs can exist for several Carrington rotations, triggering recurrent magnetic storms and consequential electron acceleration in the radiation belt \cite{Miyoshi2005,Richardson2006}. In the context of space weather, the variation of the solar wind conditions is a crucial input for models that predict the intensity of the geomagnetic activity \cite{Joselyn1995,Luo2017,Sexton2019}. The SW, as a background for coronal mass ejections (CMEs), significantly modulates the transport of CMEs by aerodynamic drag \cite{Gopalswamy2000,Subramanian2012}. Therefore, it is a fundamental task in space weather to model the speed of SW accurately.

Solar wind models can generally be grouped into two types: physical models and empirical models. Physical models take advantage of mechanisms like ballistics, Hydrodynamics (HD), and Magnetohydrodynamics (MHD) \cite{Fry2001,Odstrcil2003,Toth2005,Feng2010,Owens2020}. Empirical models aim to construct the relationship between the solar wind parameters in the interplanetary space and the observational parameters of the Sun, for example, the Wang-Sheeley (WS) model, the distance from the coronal hole boundary model (DCHB), and the Wang-Sheeley Arge (WSA) model \cite{wang1990solar,Arge2000,Riley2001,Arge2003}, whose primary parameters on the Sun are angular distances from the coronal hole boundary, expansion factors of the coronal magnetic field at the source surface, etc.

The last few years have witnessed the rise of machine learning (ML) techniques and especially neural networks (NNs). The advantages of ML neatly meet the need to build empirical models, finding embedded relationships between observational inputs and target outputs. ML has been applied in various areas of space weather, including radiation belt modeling \cite{Smirnov2020}, CME arrival time estimation \cite{Liu2018}, and geomagnetic activity indices forecasting \cite{Tan2018,Siciliano2021}. A grand review by \citeA{Camporeale2019} shows the progress of ML in space weather and points out the challenges and problems specific to space weather. As for the task of predicting the near-Earth solar wind, which people usually use the solar wind measured at the Lagrange-1 (L1) point of the Sun-Earth system to represent, there have been several successful works. Using a full-connected artificial neural network with a single hidden layer, \citeA{yang2018} developed parametric models of solar wind speed at four solar cycle phases and achieved a general performance for a randomly distributed test set with CC$=0.73$ (Correlation Coefficient, which we will define below)and RMSE$=68$km/s (Root Mean Square Error, which we will define below). The input of Yang’s models includes historical solar wind speed observation, corona number density inversed from the observed polarized brightness, and several parameters derived from the source surface magnetic field. The magnetogram of Global Oscillation Network Group (GONG) works as input of the Potential-Field Source-Surface (PFSS) algorithm \cite{Altschuler1969,Schatten1969}, to make magnetograms in the source surface ($\sim$2.5 Rs from the sun), which is helpful for a couple of other models like WSA. Considering the high correlation between the fast stream and the coronal hole, the EUV images taken by the Solar Dynamics Observatory (SDO) using the Atmospheric Image Assembly (AIA) are natural candidates for the input of the predicting models \cite{Lemen2011}, which aim to map the image data to their targets. 

In the area of machine learning, convolutional neural networks (CNN) are widely used for structured data, especially images, to extract helpful features. It was first brought up by LeCun \cite{Lecun1998} to fulfill the task of handwriting identification. Several works follow this idea. For example, \citeA{Upendran2020} combined pre-trained CNN and LSTM Recurrent Neural Network \cite{Hochreiter1997} to construct a prediction model with AIA 193 and 211 \AA images. With a cross-validation scheme, it has a performance of CC$=0.55$. We point out that the work by \citeA{yang2018} is not entirely superior to that by \citeA{Upendran2020} because the work by \citeA{yang2018} consists of four models to fit for different phases of a solar cycle, and these two works use different evaluation schemes for models, i.e., whether to use the cross-validation. After \citeA{Upendran2020}, \citeA{Raju2021} proved that simpler CNN and only AIA 193 \AA images could have a competitive performance to \citeA{Upendran2020}, and \citeA{Brown2022} used attention-based models to improve the performance, delivering an 11.1\% lower RMSE and a 17.4\% higher prediction correlation. However, we do not have an ensemble of the PFSS magnetogram and the CNN. As mentioned above, PFSS magnetograms contain rich information to predict solar wind speed. We expect CNN can promote its capability and help to find a direct mapping from the coronal magnetic field to the ambient solar wind upstream of the Earth. So in our work, we combine the CNN and the PFSS magnetograms to construct a model that forecasts the near-Earth solar wind speed. This model, being straightforward enough, can catch the necessary information in the PFSS magnetograms. It stresses the importance of the source surface magnetogram to the solar wind prediction task.

The following part of this paper consists of 3 Sections: In Section 2, we introduce our data set and methodology. We present our result and analysis of the model in Section 3. In Section 4, we present a summary and draw conclusions.

%


%
%
%
%

\section{Data and Methodology}
Our model can be described as a combination of PFSS and CNN. The PFSS model first obtains the magnetic fields at the source surface according to the photospheric magnetograms offered by GONG. Then a CNN model, with the input being the source surface magnetic fields, is trained to predict the near-Earth solar wind speed, which will be aligned and compared with the OMNI data.

\subsection{OMNI solar wind speed data}
In our work, we use the solar wind speed data from the OMNI database with a 1-hour time resolution  as the upstream/ambient solar wind speed near the Earth \cite{omniHourlyData}. The OMNI database is supported by NASA's Space Physics Data Facility (SPDF). It integrates observations of solar wind plasma and magnetic fields from different satellites located near the L1 point of the Sun-Earth system. The data are time-shifted according to the parameters, including the solar wind speed and the projected distance of displacement between the spacecraft and the Earth in the heliocentric radial direction to better represent the state of the solar wind arriving at the magnetosphere. The data are available at \url{https://omniweb.gsfc.nasa.gov}.

We use hourly OMNI solar wind speed data from September 2006 to November 2020. We eliminate the solar wind speed data corresponding to the ICME crossings according to the ICME database by \citeA{Richardson2010}.

\subsection{GONG magnetograms and the Potential Field Source Surface (PFSS) model}
As is introduced above, the photospheric magnetograms of the Global Oscillation Network Group (GONG) are commonly used magnetograms for solar wind modeling. The GONG team set up a network of six observatories on Earth for a full-day measurement of the photospheric magnetic field. GONG magnetograms have been available since 2006. The time resolution of magnetograms is four hours from September 2006 to November 2012 and one hour since November 2012. GONG provides $180^\circ\times360^\circ$ full-Carrington-longitude magnetograms that are suitable for the PFSS model.

The PFSS model is a physical model widely used to extrapolate the global coronal magnetic field\cite{Altschuler1969,Schatten1969}. The model assumes a current-free condition between the solar surface and the source surface, where the SW is supposed to start emanating along the radially-directed open field lines. It solves the Laplace equation, which is accompanied by an outer boundary condition of the equipotential surface describing the radial direction of the magnetic field and an inner boundary condition constrained by the observed photospheric magnetogram. In this work, we set the position of the source surface at 2.5 solar radii from the solar center, i.e., $R_{\rm SS}=2.5R_\odot$. We use a Python package implementing a finite difference solver called \textit{pfsspy} for the extrapolation of the global coronal magnetic field \cite{Stansby2020}.

\subsection{The convolutional neural network}
The input of the model is source surface magnetograms obtained by PFSS. Considering that the solar wind speed generally ranges from 300 to 700 kilometers per second, corresponding to a transport time to the Earth roughly from three to six days, we use four magnetograms with different days ahead of the near-Earth measurements as the input. They are magnetograms three, four, five, and six days before the prediction time, i.e., $B_{\rm SS}(t-3 {\rm\, days}),B_{\rm SS}(t-4 {\rm\, days}),B_{\rm SS}(t-5 {\rm\, days})$ and $B_{\rm SS}(t-6 {\rm\, days})$. This setting is consistent with the setting of \citeA{Upendran2020} that generates the best performance at that time of their publication. Daily sampled AIA images and daily observed solar wind speed sequence are adopted to explore the mapping relation in \citeA{Upendran2020}, while the time resolution in our work is one hour.

To our knowledge, not all information contained in one source surface magnetogram contributes to the prediction of solar wind speed at a certain place in the ecliptic plane (e.g., near the Earth). In the source surface magnetogram, the polar part that is far from the ecliptic and the area with a Carrington longitude far away from the Earth's footprint on the source surface is less correlated with the near-Earth solar wind speed, so we cut them off, only to keep a $90\times180$ sub-magnetogram from each magnetogram at the center, as the direct input to the CNN. It makes our model lighter by four times in the amount of input data and thus more efficient. Experiments show no significant difference if we use whole magnetograms, which we do not show here for simplicity.

\begin{figure}
\includegraphics[width=1.0\textwidth]{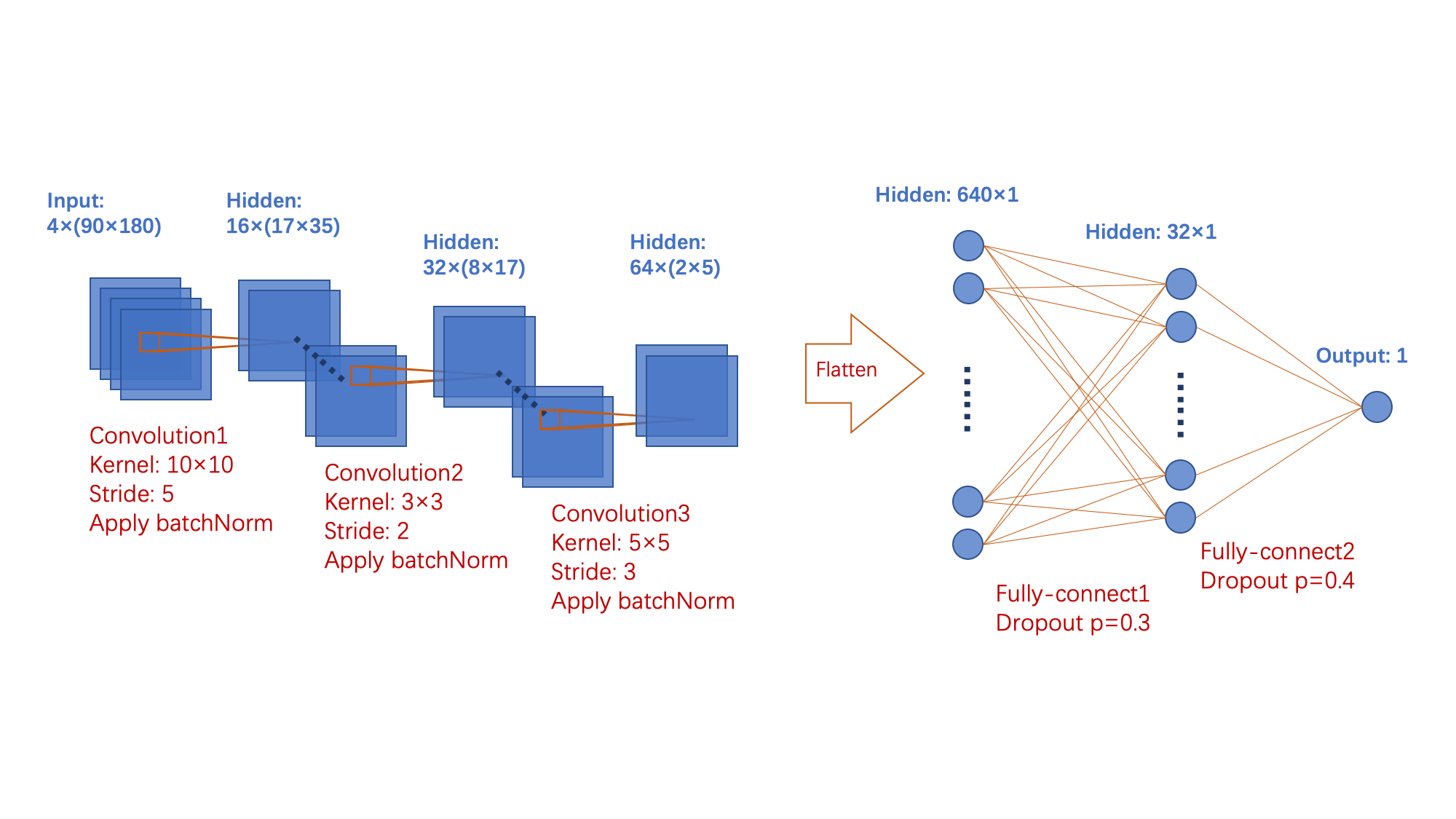}
\caption{The CNN Structure. The dark blue icons and the orange icons denote data and operations of data, respectively, and the text with the same color as data or operations are related notifications. The activation functions used for each layer are Rectified Linear Units (ReLU), except for the last layer where it is a sigmoid function.}
\label{fig:cnn_structure}
\end{figure}

\begin{figure}
\includegraphics[width=1.0\textwidth]{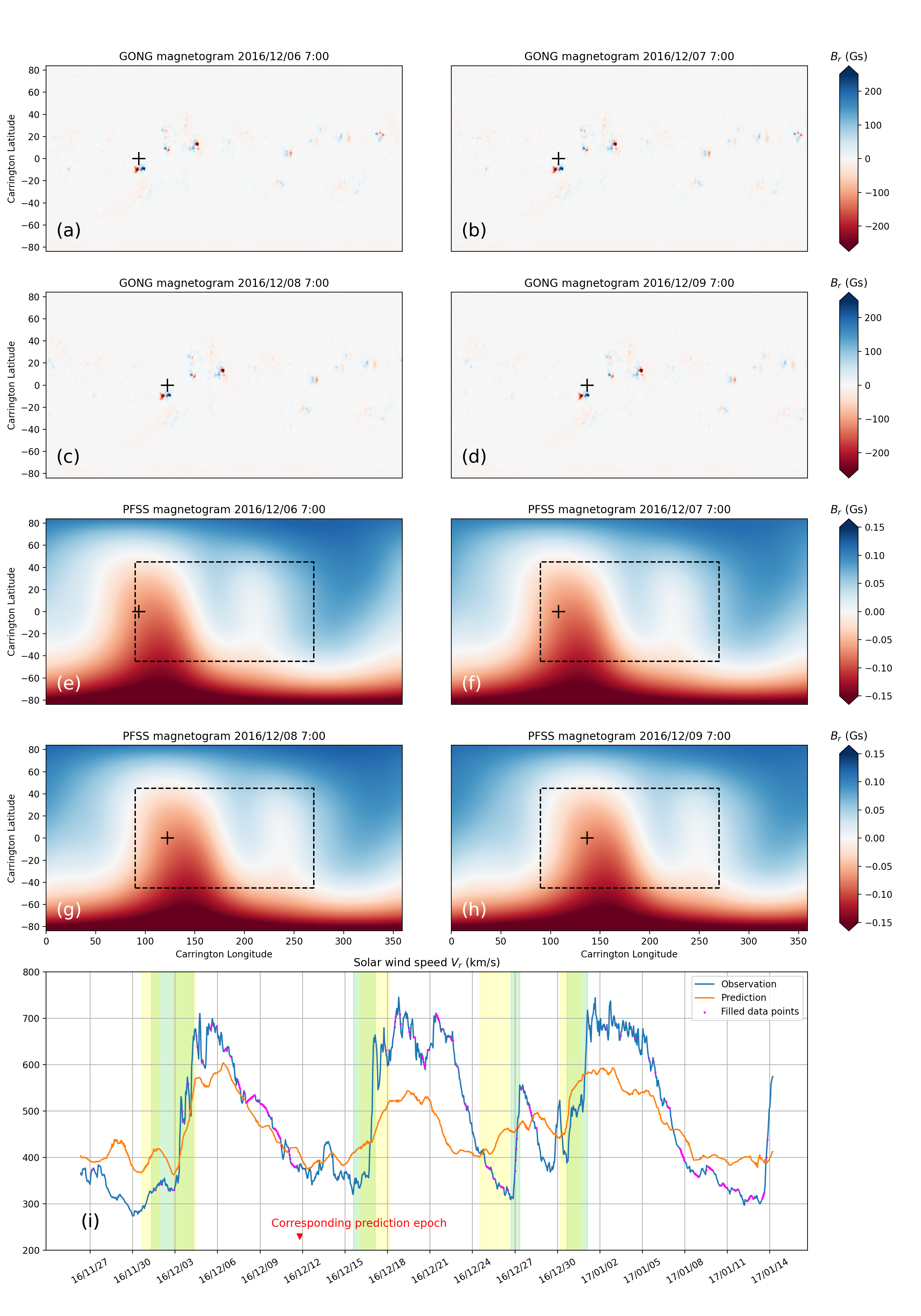}
\caption{A summary of the model. (a-d): GONG magnetograms used to obtain PFSS magnetograms. (e-h): PFSS magnetograms, i.e, $B_{\rm SS}(t-3 {\rm\, days}),B_{\rm SS}(t-4 {\rm\, days}),B_{\rm SS}(t-5 {\rm\, days})$ and $B_{\rm\, SS}(t-6 {\rm\, days})$. The black crosses in panel (a-h) are the locations of the earth in the Carrington coordinate system at the corresponding time. The black dotted rectangle denotes the reduced sub-magnetogram as the input of the CNN. (i): time series of observed and predicted solar wind speed. The red inverted triangle denotes the prediction epoch $t$ corresponding to (a-h). The epochs with a text of a certain day in the $x$-axis of (i) correspond to 12:00 UTC of that day. The green shaded areas are high speed enhancement (HSE) events (defined in Section 3) diagnosed from the observed solar wind, and the yellow shaded areas are HSE events diagnosed from the prediction. The magenta dots are linearly filled data points to the observed solar wind to achieve a time series with a one-hour cadence that can be used for the HSE diagnosis program.}
\label{fig:prediction_1}
\end{figure}

CNN is a type of artificial neural network often used in image analysis. Through two-dimensional convolution, it extracts structural features from the image data. A CNN can contain multiple convolutional layers to find more complex structures and correlations. As is displayed in Figure \ref{fig:cnn_structure}, our CNN model contains three different convolutional layers. After each convolutional layer, we apply a batch-normalization to the data \cite{Ioffe_2015_batchnorm} for more efficient convergence. After these three layers, the data go through two fully-connected layers that finally connect to a single node. The activation functions used for each layer are Rectified Linear Units (ReLU), i.e., $R(z) = \max\{0,z\}$, except for the last layer where we use a sigmoid function $S(z) = 1 / [1+\exp(-z)]$. For a better performance in generalization, we use a dropout method in fully-connected layers \cite{srivastava2014dropout}. A summary of our model is represented in Figure \ref{fig:prediction_1}, showing the original GONG data we use to obtain the input PFSS magnetogram, the input of CNN, and the sub-magnetogram we use. 

\subsection{The Training and Validating scheme}
In order to improve the performance of the model and at the same time sustain the ability to generalize, preventing over-fitting, we adopt a training and validating scheme as follows. 

We use a k-fold cross-validation scheme \cite{Dietterich1998,Upendran2020} based on quasi-continuous data segments. A quasi-continuous data segment consists of 480 continuous logs of SW speed measurements. It corresponds to 20 days of size when no ICME data is eliminated and the time resolution of logs is one hour (after the year 2012). There are 120 such segments in total. The segmentation is due to the consideration that solar wind speed data are highly time-correlated \cite{Camporeale2019}. A point-by-point random splitting, which takes every measurement of speed as an independent sampling from the statistical population, can result in similar training-, validation- and test sets because every two samples within a short time tend to be similar in a highly correlated time series. This similarity can lead to quite a welcomed training result when adopting a traditional validating scheme. However, the performance on the test set can only represent the model's ability to forecast the solar wind in a continuous segment with a size similar to the general size of continuous segments in the test set, if the test set is not continuous. In other words, point-by-point random splitting often fails when generalized to continuous predicting and thus cannot model the variation of the solar wind speed, which is not desired. After the segmentation, segments are randomly assigned into eight folds, i.e., we have $k=8$ in the k-fold cross-validation scheme. In every training cycle, we use one fold of the data as the validation set, one fold of the data as the test set, and the others as the training set. As a result, these data sets are all with randomness and continuity, suitable for generalization and a fair evaluation.

We use a gradient descent approach to train our model. The loss function applied to the training data set is a combination of the mean square error (MSE) and first-order norm of the convolutional weights. This loss term is the L-1 penalty to suspend unnecessary relations that only exist in the training data set to enhance the generalization \cite{Girosi1995}. Another method helpful for generalization is the dropout method, a common method used in neural networks. We have introduced specific dropout parameters in our model in Section 2.3. The gradient descent approach functions to minimize the loss function of the training set. At the same time, the MSEs calculated by comparing the observation and the prediction in the validating and test sets first decrease and then undulate. We save the model that generates the minimum MSE on the validating set and evaluates the model base on the performance of this model in the test set. Due to the k-fold validation, we have eight optimized models. The performance metrics are root mean square error (RMSE) and Pearson correlation coefficient (CC). They are defined as,

\begin{linenomath*}
\begin{equation}
{\rm RMSE}({\boldsymbol{x}},{\boldsymbol{y}}) = \frac{1}{n}\sqrt{\sum_{i=1}^n \left(x_{i} - y_{i}\right)^2},
\end{equation}
\end{linenomath*}
and,
\begin{linenomath*}
\begin{equation}
{\rm CC}({\boldsymbol{x}},{\boldsymbol{y}})=\frac{\sum_{i=1}^n
\left(x_i-\bar{x}\right)\left(y_i-\bar{y}\right)}
{\sqrt{\sum_{i=1}^n\left(x_i-\bar{x}\right)^2} \sqrt{\sum_{i=1}^n\left(y_i-\bar{y}\right)^2}},
\end{equation}
\end{linenomath*}
where ${\boldsymbol{x}}$ and ${\boldsymbol{y}}$ are n-dimensional vectors, and $\bar{x}$ is the arithmetic average of ${\boldsymbol{x}}$. Therefore, ${\rm RMSE}({\boldsymbol{v}}_{\rm obs},{\boldsymbol{v}}_{\rm model})$ represent general deviation of prediction, and ${\rm CC}({\boldsymbol{v}}_{\rm obs},{\boldsymbol{v}}_{\rm model})$ stands for the ability of our model to predict the variation trending.

\begin{figure}
\includegraphics[width=1.0\textwidth]{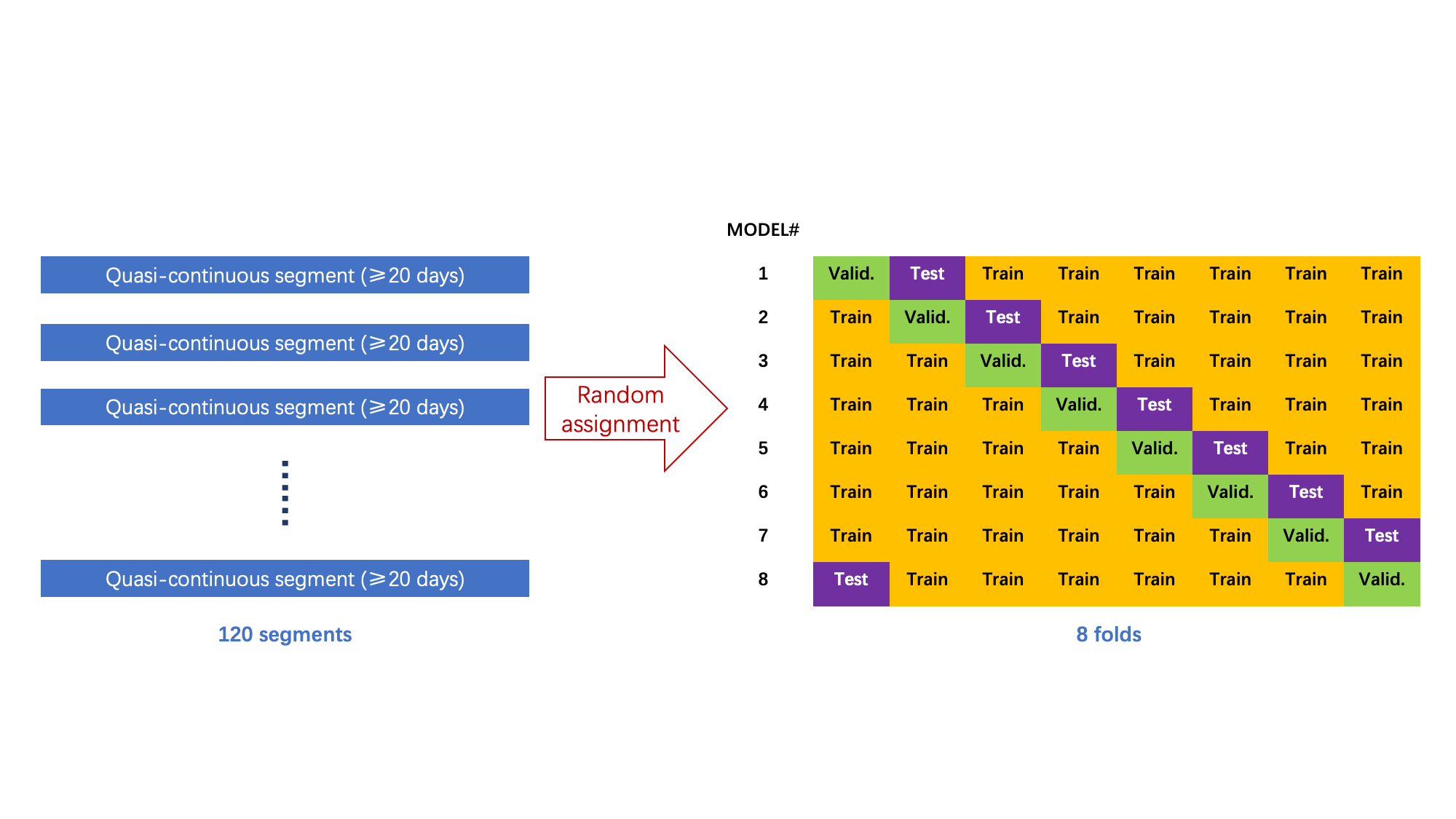}
\caption{The setting of the k-fold validation. The data are firstly split into 120 quasi-continuous segments and then randomly assigned into eight folds. The training-validating process is repeated eight times, with different folds as the training, validating, and test sets, and thus generates eight optimized models.}
\label{fig:k_fold}
\end{figure}

\begin{figure}
\includegraphics[width=1.0\textwidth]{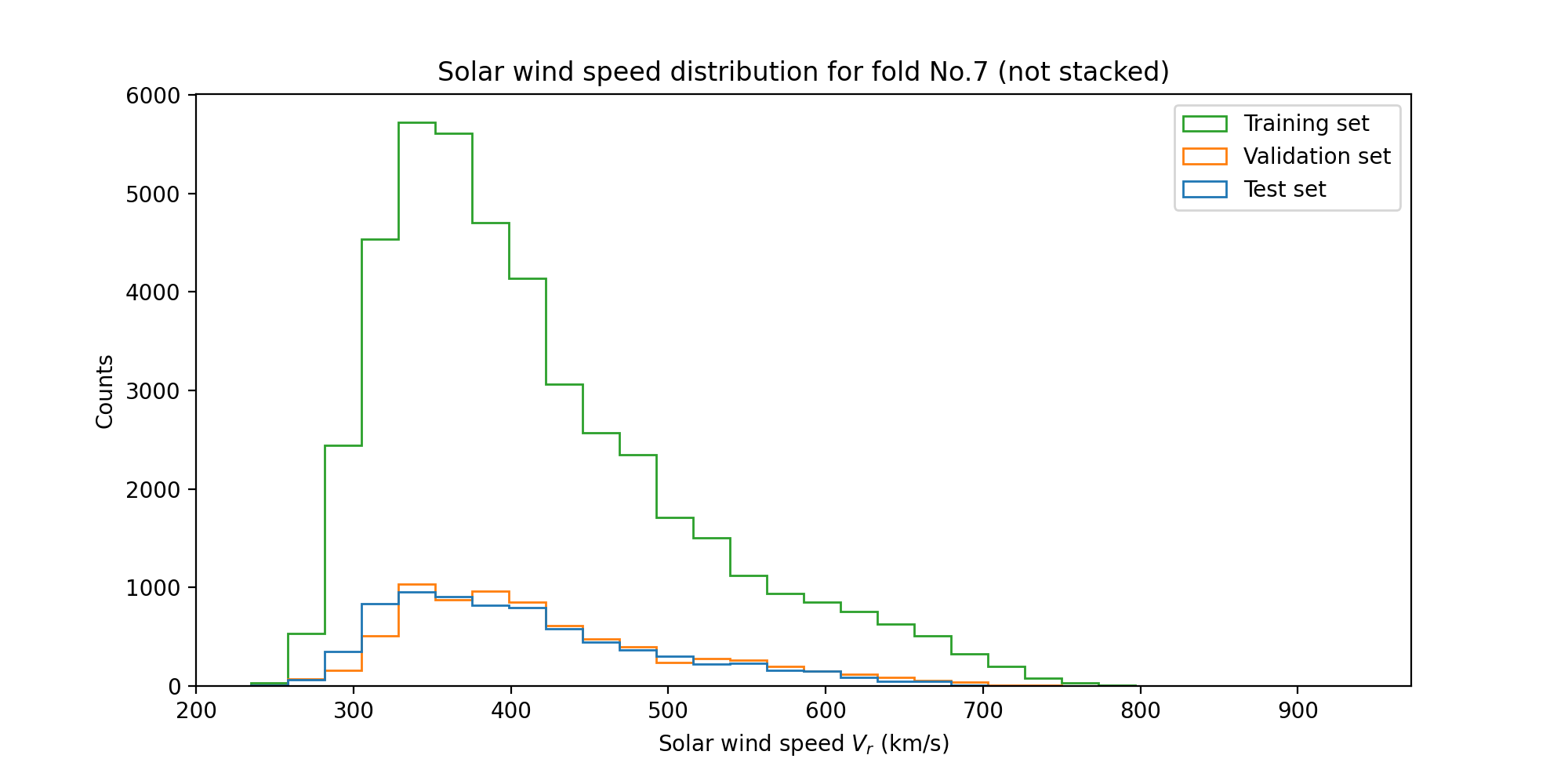}
\caption{Histograms showing distributions of the solar wind in the training (green), validation (orange), and test (blue) set of Fold No.7. These histograms are not stacked.}
\label{fig:speed_distribution}
\end{figure}

A diagram representing the k-fold validating scheme is displayed in Figure \ref{fig:k_fold}. As an example, the distributions of the solar wind in the training, validation, and test set of Fold No.7 are shown in Figure  \ref{fig:speed_distribution}.

\section{Result}

\begin{figure}
\includegraphics[width=1.0\textwidth]{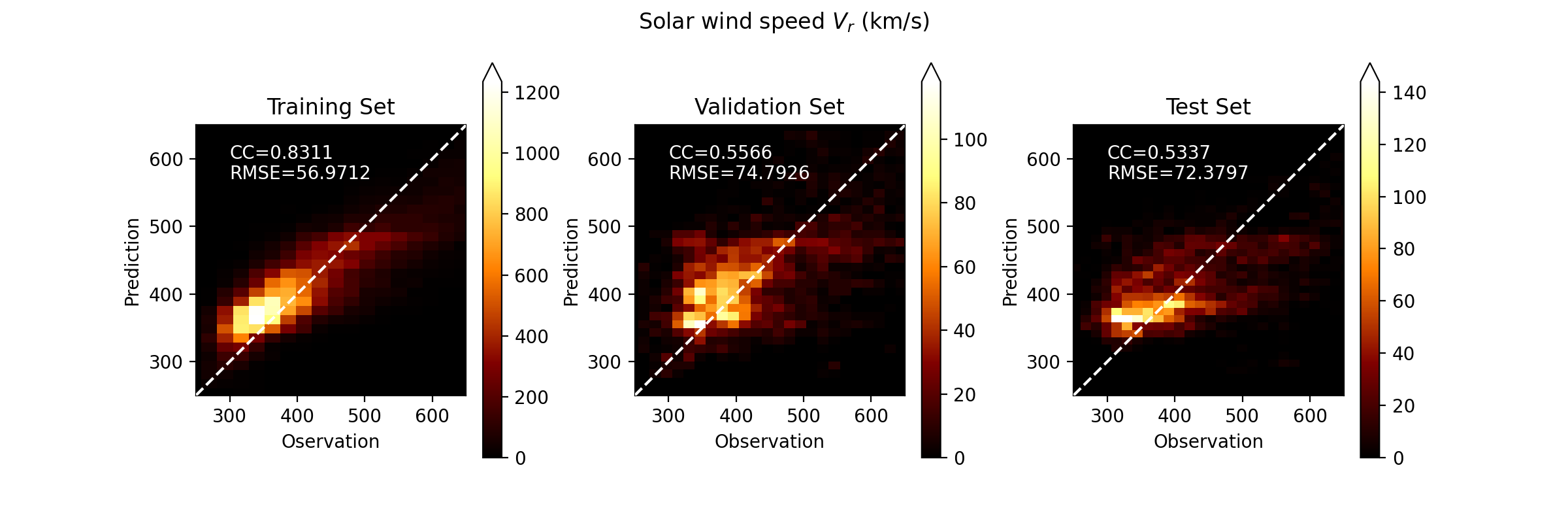}
\caption{Comparison of observations and model predictions in the training, validation, and test data set of Fold No.7.}
\label{fig:performance_on_datasets}
\end{figure}

The overall ${\rm CC}({\boldsymbol{v}}_{\rm obs},{\boldsymbol{v}}_{\rm model})$ is $0.52\pm0.06$, and ${\rm RMSE}({\boldsymbol{v}}_{\rm obs},{\boldsymbol{v}}_{\rm model})$ is $80.8\pm4.8$ km/s, averaged from the eight folds. This performance is beyond two frequently-used benchmarks, the WSA and 27-day persistence models (PS27) \cite{Owens2013}, and very close to the performance of CNN models based on AIA images, showing that PFSS magnetogram has relatively abundant information for solar wind speed forecasting and can be a candidate in ensemble forecasting networks \cite{Reiss2019}. We note here that our model has finer time resolution than the models of \citeA{Upendran2020} and \citeA{Raju2021}, whose time resolutions are one day and two hours. As a representative, the model performance with respect to Fold No.7 is shown in Figure \ref{fig:performance_on_datasets}. The comparison in the test data set indicates that our model is conservative and tends not to generate high speed streams, which may be due to the speed distribution (see in Figure \ref{fig:speed_distribution}) of the training data. In the distribution, the tail faster than 550 km/s and the tail slower than 300 km/s only take 10.7\% and 4.4\% of the whole population, respectively, which contributes less when calculate the loss function and are thus worse fitted.

Another important metric of model performance refers to the high speed enhancement (HSE) event-based evaluation. We diagnose HSE by such a criterion simplified from the criterion of \citeA{Jian2015}, and the same as \citeA{Raju2021}:
\begin{enumerate}
    \item Identify all points where speed differences are greater than 50 km/s from the previous day.
    \item Group the contiguous blocks as HSEs. Calculate the start and end time of HSEs and discard all isolated points.
    \item If the duration is less than 0.5 day, then that HSE is discarded.
    \item From the point identified as HSEs, find the minimum speed within two days prior to the start time of HSE and mark it as $V_{\rm min}$.
    \item From the start time of an HSE to one day after the HSE, find the maximum speed and mark it as $V_{\rm max}$.
    \item Find the last time reaching $V_{\rm min}$ and first time reaching $V_{\rm max}$. This duration is marked as stream interaction regions (SIRs).
    \item Regroup SIRs, find $V_{\rm min}$ and $V_{\rm max}$ for each SIR again. Eliminate redundant SIRs.
    \item SIRs with $V_{\rm min}$ greater than 500 km/s and $V_{\rm max}$ less than 400 km/s or difference between $V_{\rm max}$ and $V_{\rm min}$ less than 100 km/s are discarded.
\end{enumerate}
After identifying all HSEs in the observed time series, we identify HSEs in the predicted time series with a criterion slightly modified from above. Because the model tends to generate conservative predictions, we change the speed difference threshold in Step.1 to 35 km/s and change Step.8 as "SIRs with a difference between $V_{\rm max}$ and $V_{\rm min}$ less than 50 km/s are discarded". 

The matching status of HSEs in the observed speed sequence to HSEs in the predicted speed sequence can be grouped into three types,
\begin{enumerate}
    \item A Hit. When the start of a predicted SIR is within two days of the real start of a SIR, we mark it as a Hit.
    \item A Miss. When there is no Hit for a real SIR, we mark out a Miss.
    \item A False alarm. When there is no real SIR that can make a Hit for a predicted SIR, we mark out a False alarm.
\end{enumerate}

\begin{figure}
\includegraphics[width=1.0\textwidth]{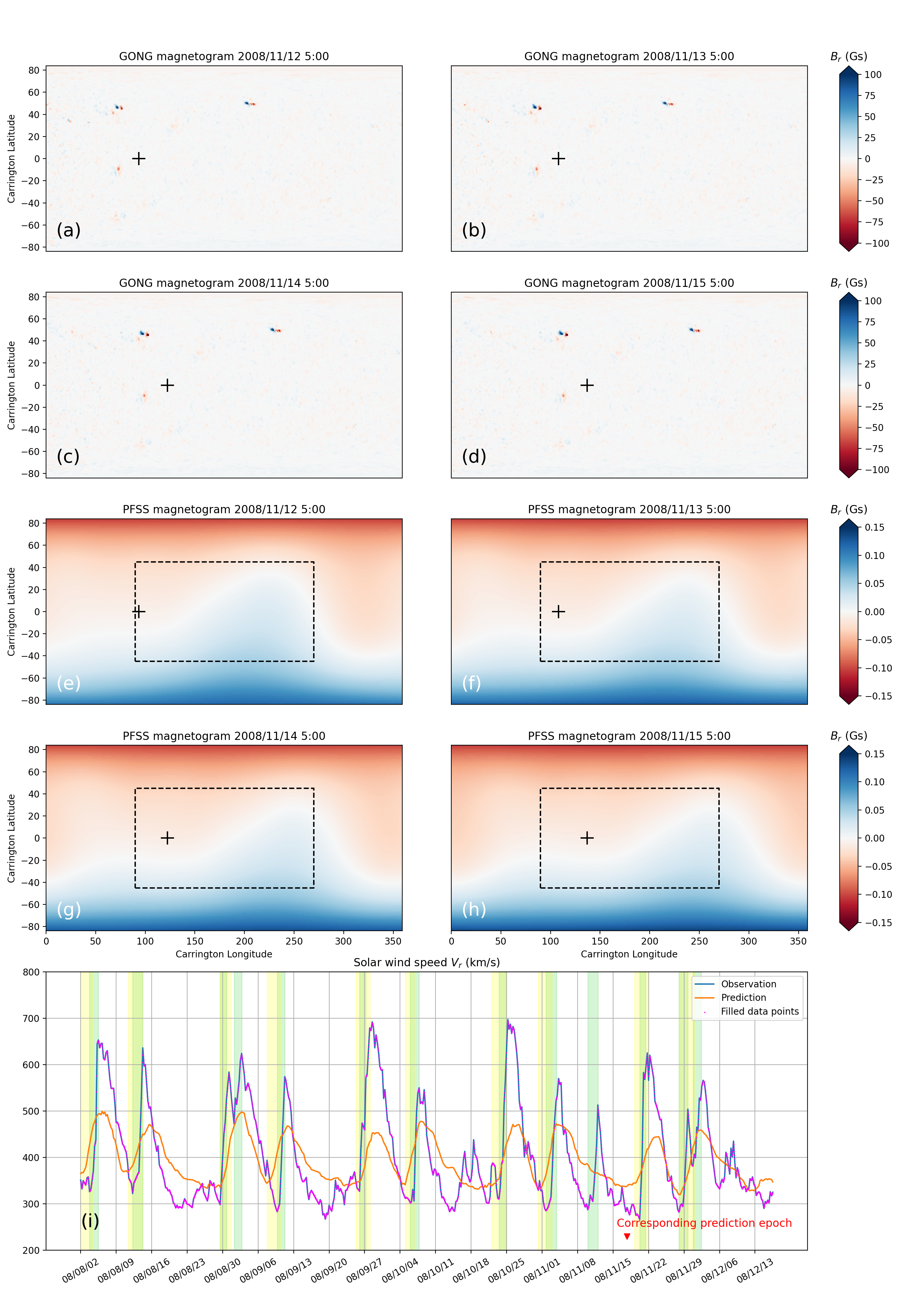}
\caption{A Similar figure to Figure \ref{fig:prediction_1} to show the model input and output for data with a time from August to December 2008 for comparison. We note here that the GONG data has a time resolution of 4 hours before November 2012, so the segment in panel (i) covers several months with the same number of logs as the segment in Figure \ref{fig:prediction_1}(i), and there are more data points that need to be filled.}
\label{fig:prediction_2}
\end{figure}

Here we present two segments of predictions and their event-based performances of them in Figure \ref{fig:prediction_1}(i) and Figure \ref{fig:prediction_2}(i). They are from test sets of Fold No.0 and Fold No.4, respectively, which have performances better than average. In these two panels, we are pleased to see a welcomed comparison result that all HSEs are Hits, although the predictions are still conservative, and the model always underestimates the peak speed. We find the models generally give HSEs predictions with SIR starts ahead of real SIR starts, but most of the start-time pairs have a difference of less than one day. This is also true for other segments and predictions (not shown here). The durations of SIRs are generally estimated as comparable with real ones, which is also true for other segments, which may be due to a balance between two factors, the relatively moderate variation the model predicts and the underestimated peak speeds. 

There are some parameters that quantify the event-based performance and give statistical evaluation over the whole data set with the number of Hits, Misses, and False alarms. They are Bias, positive predictive value (PPV), False alarm rate (FAR), and threat score (TS), which are defined as
\begin{linenomath*}
\begin{equation}
{\rm Bias} = \frac{{\rm Hits + Misses}}{\rm Hists+ False\ alarms},
\end{equation}
\end{linenomath*}

\begin{linenomath*}
\begin{equation}
{\rm PPV} = \frac{{\rm Hits}}{\rm Hists+ False\ alarms},
\end{equation}
\end{linenomath*}

\begin{linenomath*}
\begin{equation}
{\rm FAR} = \frac{{\rm False\ alarms}}{\rm Hists+ False\ alarms} = 1-{\rm PPV},
\end{equation}
\end{linenomath*}
and
\begin{linenomath*}
\begin{equation}
{\rm TS} = \frac{{\rm Hits}}{\rm Hists + Misses +False\ alarms}.
\end{equation}
\end{linenomath*}
A good model has a low Bias close to 1.0, a high PPV close to 100\%, and a high TS close to 100\%. Our model has a Bias=1.52, PPV=71\%, FAR=29\%, and TS=0.39. The TS is between the performance of the model of \citeA{Upendran2020} (TS=0.35) and the model of \cite{Raju2021} (TS=0.46), and the Bias outperforms the model of Raju's (Bias=2.15), suggesting that our model has generally comparable potential to forecast high speed stream. However, this model has a PPV that is not close enough to 100\%, so its prediction must be treated with care and combined with other methods like coronal hold area regression \cite{Bu2019} in actual business circumstances. 

\section{Summary and Discussion}
By constructing a CNN model, we prove that the map relation between the source surface magnetic field and the near-Earth solar wind speed is helpful for the forecasting task. We calculate the source surface magnetograms using the photospheric magnetic field measurement and a potential field assumption, and adopt CNN to construct a neural network architecture. We chose a series of four reduced source surface magnetograms as the input. The data pass through convolutional and full-connected layers to extract and gather structural information in the magnetograms, which is related to the information of near-Earth solar wind speed. The performance of the model is satisfying.

Our model has advantages as follows:
\begin{enumerate}
    \item The model is trained with observations from the whole solar cycle, so this trained model alone can apply to multiple phases of a solar cycle in predicting the solar wind speed from the source surface magnetic field. One can also develop multiple models for different phases in one solar cycle in specific applications, just like in the work of \citeA{yang2018}. We expect better performance for this multiple-model method.
    \item The model can deal with data with a time resolution of one hour and can predict high speed enhancement events with generally welcomed onset time and duration.
    \item The models use reduced sub-magnetograms at the source surface, which keeps the potential to develop an off-elliptic model for the prediction of high-latitude solar wind speed, once we have more off-elliptic observations.
\end{enumerate}

The main limitation of our model is being conservative due to the data distribution, but so far, people do not see a breakthrough in this area. Apart from the conservativeness, there may be other origins of forecasting bias. First, the inaccuracy of GONG observation and GONG synoptic maps. GONG uses a synthesis algorithm to estimate the magnetic field on the sun's far side, which may introduce inaccuracy. Second, the assumption of potential field corona is imperfect, and a constant $R_{\rm SS}$ of 2.5 times the solar radius may not be the reality \cite{Schulz1997}. Finally, simple CNN architecture may be less powerful than more advanced neural network settings, for example, the "transformer" that can better capture the information in images. We expect a combination of our idea and more advanced machine learning techniques in the future. An ensemble of different models that absorbs our model is also a perspective. We are more than pleased to witness that the source surface magnetic field, which contributed much to solar wind modeling in the past, can find its place in the new artificial intelligence era for space weather prediction.

\section{Open Research}
Data availability: The original data is provided by the OMNI database of NASA's Space Physics Data Facility (SPDF) and the database of Global Oscillation Network Group (GONG). For need of the analyzed data please contact the author.






\acknowledgments

The authors are grateful to the teams of OMNI and GONG for providing the data, and the teams of pfsspy for developing the program. The work at Peking University is supported by CNSA (D050106), National Key R\&D Program of China (2021YFA0718600 and 2022YFF0503800), and NSFC (42241118, 42174194, 42150105, and 42204166).

This work utilizes data from the National Solar Observatory Integrated Synoptic Program, which is operated by the Association of Universities for Research in Astronomy, under a cooperative agreement with the National Science Foundation and with additional financial support from the National Oceanic and Atmospheric Administration, the National Aeronautics and Space Administration, and the United States Air Force. The GONG network of instruments is hosted by the Big Bear Solar Observatory, High Altitude Observatory, Learmonth Solar Observatory, Udaipur Solar Observatory, Instituto de Astrofisica de Canarias, and Cerro Tololo Interamerican Observatory.


%
%



\bibliography{agusample}

%
%
%
%

\end{document}